\documentclass{aa}
\usepackage[varg]{txfonts}
\usepackage{subfigure}
\usepackage{rotating}
\usepackage{color}

\begin{document}

\title{Observations by GMRT at 323 MHz of radio-loud quasars at $z>5$}

\author{Yali Shao\inst{1, 2}\thanks{yshao@mpifr-bonn.mpg.de}
\and Jeff Wagg\inst{3}
\and Ran Wang\inst{2}
\and Chris L. Carilli\inst{4, 5}
\and Dominik A. Riechers\inst{6, 7}
\and Huib T. Intema\inst{8, 9}
\and Axel Weiss\inst{1}
\and Karl M. Menten\inst{1}}

\institute{Max-Planck-Institut f\"ur Radioastronomie, Auf dem H\"{u}gel 69, 53121 Bonn, Germany
\and Kavli Institute for Astronomy and Astrophysics, Peking University, Beijing 100871, China
\and SKA Organization, Lower Withington Macclesfield, Cheshire SK11 9DL, UK
\and National Radio Astronomy Observatory, Socorro, NM 87801-0387, USA
\and Cavendish Laboratory, University of Cambridge, 19 J. J. Thomson Avenue, Cambridge CB3 0HE, UK
\and Department of Astronomy, Cornell University, Space Sciences Building, Ithaca, NY 14853, USA
\and Max-Planck-Institut f\"ur Astronomie, K\"onigstuhl 17, D-69117 Heidelberg, Germany
\and Leiden Observatory, Leiden University, Niels Bohrweg 2, 2333 CA Leiden, The Netherlands
\and International Centre for Radio Astronomy Research, Curtin University, GPO Box U1987, Perth,
WA 6845, Australia}

\abstract{We present Giant Metrewave Radio Telescope (GMRT) 323 MHz radio continuum observations toward 13 radio-loud quasars at $z>5$, sampling the low-frequency synchrotron emission from these objects.  
Among the 12 targets successfully observed, we detected 10 above $4\sigma$ significance, while 2 remain undetected. All of the detected sources appear as point sources.
Combined with previous radio continuum detections from the literature,  9 quasars have power-law spectral energy distributions throughout the radio range; for some the flux density drops with increasing frequency while it increases for others. Two of these sources appear to have spectral turnover. 
For the power-law-like sources, the power-law indices have a positive range between 0.18 and 0.67 and a negative values between $-0.90$ and $-0.27$. 
For the turnover sources, the radio peaks around $\sim1$ and $\sim10$ GHz in the rest frame, the optically thin indices are $-0.58$ and $-0.90$, and the optically thick indices are 0.50 and 1.20. 
A magnetic field and spectral age analysis of SDSS J114657.59+403708.6 at $z=5.01$ may indicate that the turnover is not caused by synchrotron self-absorption, but rather by free-free absorption by the high-density medium in the nuclear region. Alternatively, the apparent turnover may be an artifact of source variability.
Finally, we calculated the radio loudness $R_{2500\rm\, \AA}$ for our sample, which spans a very wide range from 12$^{+13}_{-13}$ to 4982$^{+279}_{-254}$. }


\maketitle

\section{Introduction} \label{intro}

High-redshift quasars at $z>5$ are critical for probing the physical conditions at the end of the reionization epoch, and for studying the early evolutionary stage of  the formation of active galactic nuclei (AGN) and the possible coevolution of the first supermassive black holes (SMBHs) and their host galaxies. 
A significant number of quasars have been discovered at $z>5$ (e.g., \citealt{Fan2006lala}; \citealt{Jiang2015}; \citealt{Mazzucchelli2017}). However, only a few of these are classified as radio-loud quasars (e.g.,  \citealt{Anderson2001};  \citealt{Fan2001}; \citealt{Sharp2001}; \citealt{Romani2004}; \citealt{McGreer2006}; \citealt{McGreer2009}; \citealt{Willott2010}; \citealt{Zeimann2011}; \citealt{Yi2014}; \citealt{Banados2018}; \citealt{Belladitta2020}).  
High-redshift radio-loud quasars are interesting because their prominent radio jets should directly relate to the activity of their central SMBHs.

Previous studies of these high-redshift radio-loud quasars at $z>5$ focus on their high-frequency properties through GHz-observations from 1.4 GHz to 91 GHz (e.g., \citealt{Romani2004}; \citealt{Frey2005}; \citealt{Momjian2008}; \citealt{Sbarrato2013}; \citealt{Cao2014}; \citealt{An2020}). 
Ten of these have been observed with milli-arcsecond (mas) resolution by the European Very Long Baseline Interferometry (VLBI) Network (EVN) or the Very Long Baseline Array (VLBA). Seven out of ten show single-compact or double-structure morphology (\citealt{Frey2003,Frey2005,Frey2008,Frey2011};  \citealt{Momjian2003, Momjian2018}; \citealt{Cao2014}; \citealt{Gabanyi2015}), which together with the steep radio spectrum makes these quasars similar to compact symmetric objects (CSOs). Of these objects, 30$\%$ are identified as blazars, which have a core-jet morphology (\citealt{Romani2004}; \citealt{Frey2010,Frey2015}).
The following three blazars have been studied from radio to X-ray wavelengths, and even to $\gamma$-rays: SDSS J114657.79+403708.6 (hereafter J1146+4037) at $z=5.01$, identified by \citet{Ghisellini2014}; SDSS J102623.61+254259.5 (hereafter J1026+2542) at $z=5.3$, identified by \citet{Sbarrato2012,Sbarrato2013}; and Q0906+6930 at $z=5.47$, identified by \citet{Romani2004}.  
Through multi-epoch EVN VLBI observations toward J1026+2542 at $z=5.3$, \citet{Frey2015} detected a significant structural change of the jet and, therefore, estimated the apparent proper motion of three components of the jet. 
However, there are few observations of these quasars below 1 GHz, except for information from a handful of  low-frequency sky surveys such as the LOFAR Two-metre Sky Survey (LoTSS; \citealt{Shimwell2017}), the TIFR GMRT Sky Survey (TGSS; \citealt{Intema2017}), and the GaLactic and Extragalactic All-sky Murchison Widefield Array (GLEAM; \citealt{Hurley-Walker2017}) survey. 
In this paper, we extend the  radio information of these high-redshift radio-loud quasars at $z>5$ to lower frequencies through Giant Metrewave Radio Telescope (GMRT) observations at 323 MHz.

One of the outstanding problems in the study of AGNs is understanding the origin and evolution of the radio emission. The merger or accretion events in the host galaxies can transfer fuel to feed the central black holes (BHs), which are believed to give rise to radio activity (e.g., \citealt{Chiaberge2011}). 
As a consequence, the radio emission evolves in a dense and possibly inhomogeneous ambient medium, which can influence BH growth at least at the early stages. The ideal targets for better understanding this  problem are young radio sources, whose radio lobes still reside within the innermost region of the host galaxy. 

Compact symmetric objects are very powerful ($P_{1.4 \rm\, GHz}> 10^{25}$ W Hz$^{-1}$), compact (linear size of 50$-$100 pc), and young ($\lesssim10^{4}$ yr) objects with a rather symmetric radio structure and convex synchrotron radio spectra (\citealt{Wilkinson1994}; \citealt{Murgia2003}; \citealt{Polatidis2003}). The characterized convex synchrotron radio spectrum peaks at around 100 MHz in the case of compact steep spectrum (CSS) sources, at about 1 GHz in the case of GHz-peaked spectrum (GPS) objects, and up to a few GHz (e.g., 5 GHz) in the case of high-frequency peakers (HFPs; \citealt{Dallacasa2000}). The turnover is explained as synchrotron self-absorption (SSA) affecting in a small radio-emitting region or the free-free absorption (FFA) by the dense ambient medium. In a ``young scenario'', as the source grows, the inner region (possibly a tiny radio lobe) expands, and as a result, the turnover frequency moves to lower frequencies. In this scenario, the HFPs are newborn radio sources that develop into extended radio sources (e.g., FR I, FR II) after evolving through GPS and CSS stages. It is possible that the activity is recurrent in at least some sources: there have been observations of faint extended emission around a few GPS sources (e.g., \citealt{Baum1990}; \citealt{Stanghellini1990}; \citealt{Marecki2003}). The extended emission could be a relic of an earlier active period, into which the reborn radio jets are expanding. Another popular explanation for the turnover and compact natures of CSS, GPS, and HFP is the ``frustration'' hypothesis. This theory argues that these sources are confined in small spatial scale and high-density environments, and as a consequence the radio emission is frustrated by the abundant nuclear plasma (\citealt{vanBreugel1984}; \citealt{Peck1999}; \citealt{Tingay2003}; \citealt{Callingham2015}; \citealt{Tingay2015}). In addition, \citet{An2012} also proposed that young sources with strong constant AGN power breaking through the dense inner region of the host galaxy could result
in the compact morphology and the turnover properties of CSOs.   

However, searching for and investigating the onset of radio AGNs through CSOs, especially GPS/HFPs at low redshifts, requires high-frequency (e.g., $>5$ GHz for HFPs) observations. In contrast, considering that for high redshifts, the spectral turnover shifts to lower frequencies, high-redshift radio-loud quasars make a unique sample for low-frequency radio studies.

In this paper, we report on 323 MHz GMRT observations toward 13 radio-loud quasars at $z>5$ carried out to investigate their low-frequency synchrotron properties and better sample their radio spectra. 
This paper is organized as follows. 
In Section \ref{obs} we describe our sample, the GMRT observations, and the data reduction. 
In section \ref{res} we summarize previous studies of each target in our sample and present the new GMRT measurements. 
In Section \ref{ana},  
we apply different spectral models to the observed radio spectra, and investigate the possible origin of the spectral turnover considering magnetic field strength and spectral age for J1146+4037. In addition, we also present the radio loudness of our sample in the conventional definition with predicted 5 GHz flux density from our spectral models.
Finally in Section \ref{sum}, we present a short summary. 
Throughout this work we assume a $\Lambda$CDM cosmology with $H_{0}$ = 71 km s$^{-1}$ Mpc$^{-1}$, $\Omega_{\rm M}$ = 0.27 and $\Omega_{\Lambda}$ = 0.73 \citep{Spergel2007}. 

\section{Observations and data reduction} 
\label{obs}

Our sample consists of 13 radio-loud quasars at $z>5$.
Ten of these are known to be the brightest 1.4 GHz quasars at the highest redshifts.  We identified three of our targets  by cross-matching optical wavelength quasar catalogs with the Very Large Array (VLA) Faint Images of the Radio Sky at Twenty-Centimeters (FIRST; \citealt{Helfand2015}) catalog and the VLA high-resolution radio survey of SDSS strip 82 \citep{Hodge2011} catalog: SDSS J161425.13+464028.9 (hereafter J1614+4640), SDSS J223907.56+003022.6 (hereafter J2239+0030), and WFS J224524.20+002414.0 (hereafter J2245+0024).
We carried out GMRT 323 MHz radio continuum observations  
to study the low-frequency synchrotron properties and better sample their radio spectra to understand the evolutionary properties of the first quasar systems 
approaching the end of the cosmic reionization. 
Table \ref{pdata1} and \ref{pdata2} present our sources summarizing the  archival radio data.

\subsection{Observations at 323 MHz by GMRT}

The 323 MHz GMRT observations were carried out 2015 November 9 and 10 (proposal code: 29$\_$046; PI: Jeff Wagg). 
We first observed a flux calibrator - 3C48 for 15 minutes, 
followed by 5$-$20 minutes observations on a bright phase calibrator within $10\degr$ of our targets. 
Then we performed 20$-$50 minutes observations on the target, after which we observed the phase calibrator again, followed by another 20$-$50 minutes observations of the next target before one final observation of the phase calibrator. The adopted phase calibrators are list in Table \ref{GMRTm}.
We recorded full polarization with a total band width of 32 MHz. 
The theoretical synthesized beam size is $9\arcsec$ and the observed one for each target is listed in Table \ref{GMRTm}. 

\subsection{Data reduction}
The data reduction was conducted with the package - {\ttfamily{Source Peeling and Atmospheric Modeling}} ({\ttfamily{SPAM}}\footnote{\url{http://www.intema.nl/doku.php?id=huibintemaspam}}; \citealt{Intema2014a, Intema2014b}). 
It is an {\ttfamily{Astronomical Image Processing System}} ({\ttfamily{AIPS}}\footnote{\url{http://www.aips.nrao.edu/index.shtml}}; \citealt{Greisen2003}) extension based on {\ttfamily{Python}}, 
which can work on high-resolution, low-frequency radio interferometric data efficiently and systematically. 
We followed the standard {\ttfamily{SPAM}} pipeline, which contains direction-dependent calibration, modeling and imaging for correcting mainly ionospheric dispersive delay (the detail can be seen in \citealt{Intema2017}).

\section{Results} \label{res}

For 12 of our radio-loud quasars at $z>5$ we obtained useful data. However, the data for 1 source (i.e., SDSS J013127.34$-$032100.1; hereafter J0131$-$0321) is contaminated by serious radio frequency interference (RFI). 
Ten of 12 are detected above $4\sigma$, while the remaining 2 are not detected, and we present their  3$\sigma$ upper limits. 
The GMRT 323 MHz measurements and images are presented in Table \ref{GMRTm} and Figure \ref{GMRT323}. We note that we only present formal statistical errors from Gaussian fitting  in Table \ref{GMRTm}. In our analysis (Section \ref{ana}), we also consider another $10\%$ calibration errors \citep{Chandra2004} for our 323 MHz GMRT data.

$\textbf{J0131$-$0321}$ - 
This is an optically luminous radio-loud quasar at $z=5.18$ discovered by \citet{Yi2014} using the Lijiang 2.4 m and Magellan telescopes. 
These authors measured the BH mass to be $2.7\times10^9$ $M_{\odot}$ with an $i$-band magnitude of 18.47 mag.
It is the second most luminous object in our sample with 1.4 GHz flux density of $33.69\pm0.12$ mJy from the VLA FIRST (\citealt{Helfand2015}; observation date: 2009 March 21; angular resolution: $\sim5\farcs4$).  
\citet{Ghisellini2015} conducted a spectral energy distribution (SED) fit with an accretion disk, a torus, and a jet to constrain the viewing angle $\theta$ of the jet, which is very close to the line of sight (e.g., $\theta\sim$ 3$-$5$\degr$). 
\citet{Gabanyi2015} observed this target on 2014 December 2 using EVN VLBI at 1.7 GHz. 
They found a single compact radio component with a flux density of $64.40\pm0.30$ mJy,  and estimated a relatively high brightness temperature ($T_{\rm B}$ $\ga$ $2.8\times10^{11}$ K) and a moderate Doppler boosting factor ($\ga 6$). 
In addition, these authors noted a significant flux density variation by comparing the FIRST 1.4 GHz and VLBI 1.7 GHz measurements, which together suggest that this object may be a blazar.

\begin{figure*}[h]
\centering
\subfigure{\includegraphics[scale=0.1]{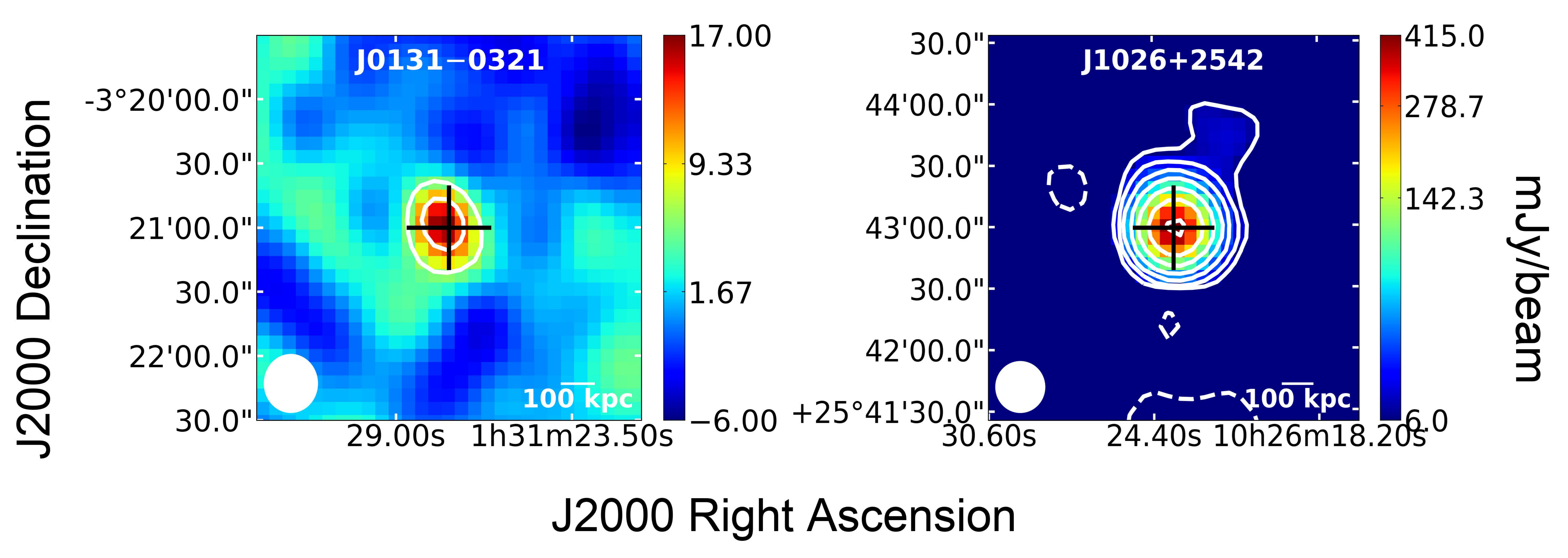}} 
\caption{Continuum map from GMRT 147.5 GHz extracted from the TGSS database \citep{Intema2017}. The black crosses show the optical positions of the quasars from \citet{Yi2014} and the SDSS. The shapes of the synthesized beams are plotted in the bottom left of each panel: J0131$-$0321 with a clean beam size of $27\farcs25\times25\farcs00$, and J1026+2542 with a synthesis beam size of $25\farcs00\times25\farcs00$. Contour levels for each map are as follows: J0131$-$0321 - $[3, 6]\times2.0$ mJy beam$^{-1}$, J1026+2542 - $[-3, 3, 6, 12, 24, 48, 96, 192]\times2.0$ mJy beam$^{-1}$. Source information is given in Table \ref{pdata1} and \ref{pdata2}.}
\label{tgss150}
\end{figure*}

Our new GMRT observations failed owing to severe RFI. 
However, we found an obvious signal $> 5\sigma$ in the TGSS \citep{Intema2017} database at 150 MHz, which is not shown in the official catalog because of their $7\sigma$ selection criterion.  
We presented the 150 MHz continuum image in the left panel of Figure \ref{tgss150},  which was observed on 2016 March 15. It appears as a point source. We measured the flux density with the {\ttfamily{Common Astronomy Software Applications (CASA\footnote{\url{https://casa.nrao.edu/}})}} 2D Gaussian tool, which shows a flux density of $25.70\pm4.70$ mJy. This is consistent with that measured by the {\ttfamily{Python Blob Detector and Source Finder}} ({\ttfamily{PyBDSF\footnote{\url{http://www.astron.nl/citt/pybdsf/}}}}).

$\textbf{SDSS J074154.72+252029.6 (hereafter J0741+2520)}$ - 
\citet{McGreer2009} discovered this target by combining data from the FIRST radio survey and the SDSS. 
The $i$-band magnitude is 18.45 mag at $z=5.194$. 

We detected the 323 MHz radio continuum emission for this target with a peak flux density of $1.73\pm0.09$ mJy beam$^{-1}$. It is an unresolved target.

$\textbf{SDSS J083643.85+005453.3 (hereafter J0836+0054)}$ - 
\citet{Fan2001} selected this target from multicolor imaging data of SDSS. 
It has a strong and broad yet partially absorbed  Ly$\alpha$ emission line, which indicates a redshift of 5.82. 
\citet{Stern2003} observed this target using FLAMINGOS on the Gemini-South 8 m telescope and calculated a slightly lower redshift of 5.774 from broad emission lines of \ion{C}{iv} $\lambda$1549 $\rm\AA$ and \ion{C}{iii}] $\lambda$1909 $\rm\AA$. 
The 1.4 GHz flux density is $1.11\pm0.15$ mJy from FIRST (\citealt{Helfand2015};  observation date: 1998 August; angular resolution: $\sim5\farcs4$), $2.5\pm0.5$ mJy from NVSS (\citealt{Condon1998};  observation date: 1993 November 15; angular resolution: $\sim45\arcsec$), $1.75\pm0.04$ mJy from VLA A-configuration (\citealt{Petric2003};  observation date: 2002 March 6; angular resolution: $\sim1\farcs5$), and $1.96\pm0.31$mJy from VLA B-configuration (\citealt{Frey2005};  observation date: 2003 October 5; angular resolution: $\sim6\arcsec$). This source shows significant variability through $\sim10$ years of monitoring.
\citet{Frey2003} carried out  VLBI 1.6 GHz observations at $\sim10$ mas angular resolution and found this target to be compact.
\citet{Frey2005} also observed this source at mas resolution with the VLBI Network at 5 GHz, which confirms that it is a compact source with a flux density of $0.34\pm0.04$ mJy. In addition, these authors also conducted an almost simultaneous VLA B-configuration, arcsecond resolution observations at 5 GHz ($f_{\nu}=0.43\pm0.06$ mJy) toward this target. By comparing the 5 GHz radio emission from observations at two different resolutions, the authors constrained the radio emission to arise from within the central 40 pc.

This target is unresolved by our 323 MHz GMRT observations and has a peak flux density of $1.94\pm0.17$ mJy beam$^{-1}$. The low surface brightness $6\sigma$ tail-like structure showing in the Figure \ref{GMRT323} of this target may be due to a low-redshift radio source in the foreground at  a projected distance of $\sim10\arcsec$, which is resolved by \citet{Frey2005}.

$\textbf{SDSS J091316.56+591921.5 (hereafter J0913+5919)}$ - 
\citet{Anderson2001} reported this target to have a relatively narrow Ly$\alpha$ emission line at a redshift of $z = 5.11$. 
The redshift measurement was improved to 5.1224 by \citet{Hewett2010} through investigating the empirical relationships between redshifts and multiple line emissions. 
The high-resolution 1.4 GHz VLBA observations \citep{Momjian2003} revealed a compact radio property.
Motivated by the compact nature and the narrow Ly$\alpha$ line, \citet{Carilli2007} searched for the redshifted \ion{H}{i} 21 cm absorption line to this object with GMRT, but unfortunately found nothing.

This object is a point source in our new GMRT observations. 
The 323 MHz peak flux density is $9.80\pm0.12$ mJy beam$^{-1}$, which makes it the second strongest detection in our  sample.

$\textbf{J1026+2542}$ - 
This quasar was selected from the SDSS catalog with a redshift of 5.28. 
It is the strongest radio-loud quasar in our sample. 
The flux density at 1.4 GHz from VLA FIRST (\citealt{Helfand2015};  observation date: 1995 November 22; angular resolution: $\sim5\farcs4$) is $239.44\pm0.14$ mJy. 
The VLA NVSS lists a similar $S_{\rm 1.4\, GHz} = 256.9\pm7.7$ mJy (\citealt{Condon1998};  observation date: 1993 December 6; angular resolution: $\sim45\arcsec$). 
\citet{Sbarrato2012, Sbarrato2013} proposed that it is a blazar by constructing a full SED analysis in the radio up to $\gamma$-ray regimes.  These authors yielded a small jet viewing angle of $\sim3\degr$ with respect to the line of sight,  and significant Doppler boosting with a large bulk Lorentz factor $\sim13$.
\citet{Frey2015}, for this object, obtained the first directly jet proper motions (up to $\sim$ 0.1 mas yr$^{-1}$) in $z > 5$ blazars with  EVN VLBI dual-frequency (1.7 and 5 GHz) observations combined with VLBA imaging by \citet{Helmboldt2007}.  The lower limit  \citet{Frey2015} obtained for the core brightness temperature, $\sim2.3\times10^{12}$ K, is consistent with the blazar nature. 

This quasar is unresolved in the new GMRT 323 MHz observations with the highest intensity among our sample with peak flux density of $317.72\pm0.33$ mJy beam$^{-1}$. We also searched the GMRT TGSS data \citep{Intema2017} and present the 150 MHz image in the right panel of Figure \ref{tgss150}; this observation took place on 2016 March 15. In addition, we found 20 separate flux density measurements across 72$-$231 MHz from the GLEAM \citep{Hurley-Walker2017} survey, which are listed in Table \ref{pdata2}.

$\textbf{SDSS J103418.65+203300.2 (hereafter J1034+2033)}$ - 
SDSS discovered this quasar at a redshift of 5.0150. 

This object is unresolved in our new 323 MHz GMRT observations with a peak flux density of $3.03\pm0.11$ mJy beam$^{-1}$.

$\textbf{J1146+4037}$ - 
This quasar is selected from the SDSS catalog with a redshift of 5.009, which has been improved to  5.0059 by \citet{Hewett2010}.
\citet{Frey2010} observed this target with mas angular resolution using VLBI at 1.6 and 5 GHz, and found that it is a compact source.

This quasar is a point source in our 323 MHz observations with $S_{\rm 323\, MHz,\,peak}$ = $3.05\pm0.07$ mJy beam$^{-1}$.

$\textbf{FIRST J1427385+331241 (hereafter J1427+3312)}$ - 
\citet{McGreer2006} discovered this quasar by matching sources from the FLAMINGOS Extragalactic Survey (FLAMEX) IR survey to the FIRST survey radio sources with NDWFS counterparts. 
It is a broad absorption line quasar with a redshift of 6.12, which makes it the second highest redshift quasar in our sample.
\citet{Momjian2008} presented 1.4 GHz VLBA observations, and found two mas-scale resolved components with a projected distance of  31 mas (174 pc). These authors assumed a CSO model that reveals a kinematic age of $\sim$ 10$^{3}$ yr.
\citet{Frey2008} studied the compact radio structure of this target with a resolution of a few mas with EVN VLBI at 1.6 GHz and 5 GHz. They also detected the double structure with a separation of 28.3 mas (160 pc), which confirms that it is a young CSO ($\leq10^{4}$ yr).

By fitting the 323 MHz GMRT image of this target, the {\ttfamily{CASA}} 2D Gaussian fit tool gives a deconvolved source size of $(7\farcs89\pm2\farcs41)\times(1\farcs85\pm0\farcs49)$, an integrated flux density of $4.51\pm0.22$ mJy, and a peak flux density of $3.96\pm0.10$ mJy beam$^{-1}$. The fact that the deconvolved major axis lies roughly along the synthesized beam major axis makes this size determination questionable and in the following we treat this target as a point source. This target is included in the VLA 325 MHz image of the Bootes field \citep{Coppejans2015}, with $S_{325\, \rm MHz} = 4.80\pm0.80$ mJy, which is consistent with our 323 MHz measurement.

$\textbf{CFHQS J142952.17+544717.6 (hereafter J1429+5447)}$ - 
\citet{Willott2010} discovered this quasar  in the Canada-France-Hawaii Telescope Legacy Survey (CFHTLS) Wide W3 region. 
This object has strong continuum but a weak Ly$\alpha$ emission line at $z$ = 6.21, meaning that  it the most distant object in our sample.
\citet{Wang2011}  detected both CO (2$-$1) line  and its underling continuum. The CO line emission shows two prominent peaks that are separated with 1$\farcs$2 (6.9 kpc). A Gaussian fit to the CO spectrum yielded a source redshift of 6.1831.
\citet{Frey2011} detected a single dominant component at both 1.4 GHz and 5 GHz by VLBI toward this target. It was slightly resolved with mas-scale resolution within $<$100 pc. The derived brightness temperature on the order of $10^{8}$ K, which supports the AGN origin of the radio emission of this quasar.

This target is a point source with a peak flux density of $4.91\pm0.16$ mJy beam$^{-1}$ in our 323 MHz observations from the GMRT.

$\textbf{J1614+4640}$ - 
This target is an SDSS quasar with a redshift of 5.31.

J1614+4640 is unresolved in our 323 MHz GMRT observations with a peak flux density of $0.94\pm0.09$ mJy beam$^{-1}$.

$\textbf{SDSS J222843.54+011032.2 (hereafter J2228+0110)}$ - 
\citet{Zeimann2011} discovered this radio-loud quasar by matching optical detections of the deep SDSS Stripe 82 with their radio counterparts in the Stripe 82 VLA Survey. 
It is an optically faint but radio bright quasar at $z=5.95$.

It is too faint to be detected in our 323 MHz GMRT observations. The resulting image  root mean square (rms) noise level is 0.15 mJy beam$^{-1}$, and we adopt 3$\sigma$ rms as the upper limit in our radio spectrum analysis.

$\textbf{J2239+0030}$ - 
This target comes from the SDSS catalog. 

This object is a point source in our GMRT survey with a peak flux density of $1.55\pm0.20$ mJy beam$^{-1}$.

$\textbf{J2245+0024}$ - 
\citet{Sharp2001} discovered this target in the Public Isaac Newton Group Wide Field Survey with a redshift of 5.17. 
This object has a strong and highly irregular Ly$\alpha$ emission line profile. 

We did not detect this target in our 323 MHz GMRT observations. The final image rms noise level is 0.12 mJy beam$^{-1}$, and we consider 3$\sigma$ rms as the upper limit in our radio spectrum analysis.

\section{Analysis and discussion} \label{ana}

\subsection{Radio spectral modeling}

To describe the radio spectrum properties of our sample, we adopt two different spectral models in this paper. For our fitting, we used the {\ttfamily{emcee}}\footnote{\url{http://dfm.io/emcee/current/}} package \citep{emcee2013} in the case of more than two data points and a least-squares method in the case of only two data points.

Firstly, we used the standard nonthermal power-law model
\begin{equation}
\label{pl}
S_{\nu} = a\nu ^{\alpha_{\rm pl}},
\end{equation}
where $a$ represents the amplitude of the synchrotron spectrum, $\alpha_{\rm pl}$ shows the synchrotron spectral index, and $S_{\nu}$ is the flux density at frequency $\nu$, in MHz.

In addition, the following generic curved model was used to characterize the entire spectrum of a peaked-spectrum source:
\begin{equation}
\label{gcm}
S_{\nu} = \frac{S_{\rm p}}{(1 - e^{-1})} (1 - e^{-(\nu/\nu_{\rm p})^{\alpha_{\rm thin} - \alpha_{\rm thick}}}) (\frac{\nu}{\nu_{\rm p}})^{\alpha_{\rm thick}},
\end{equation}
where $\alpha_{\rm thick}$ and $\alpha_{\rm thin}$ are the spectral indices in the optically thick and optically thin parts of the spectrum, respectively. The quantity $S_{\rm p}$ is the flux density at the peak frequency $\nu_{\rm p}$ \citep{Snellen1998}. This equation only performs a fit to the spectra on regions at the lower and the higher sides of the peak, but cannot discriminate between the underlying absorption mechanism (e.g., SSA or FFA) causing the spectral turnover.

\begin{figure*}
\centering
\subfigure{\includegraphics[scale=0.12]{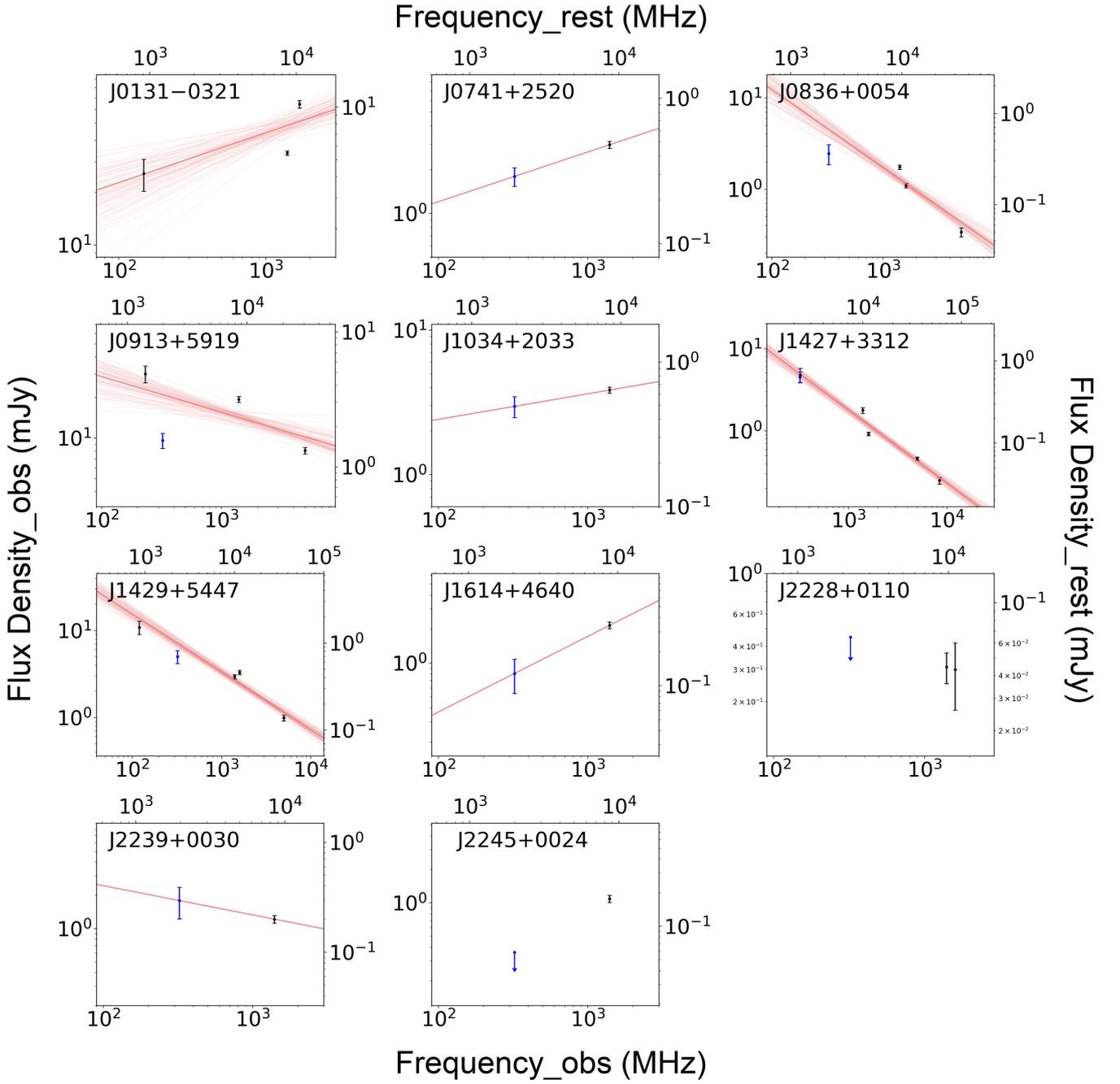}} 
\caption{Spectral model fit with the standard nonthermal power-law model. The black points with error bars represent archival data taken from the literature. The blue points with error bars or downward arrows indicate GMRT 323 MHz measurements from our work. Calibration errors of  10$\%$  were added to the statistical errors of the GMRT 323 MHz data points \citep{Chandra2004}. The red lines represent the fitted models. For fits employing the MCMC method, 100 models were randomly selected from the parameter space as red shaded areas to visualize the model uncertainties. There is not enough data in the cases of J2228+0110 and J2245+0024, thus the measurements are given without any fitting. The fitted results are presented in Table \ref{fitr}.}
\label{all}
\end{figure*}

Nine sources in this study show no clear evidence for spectral curvature (see Figure \ref{all}), so we only  fit the nonthermal power-law model and present the fitted results in Table \ref{fitr}. Four of 9 show increasing flux density with increasing frequency and have power-law indices between 0.18 and 0.67. The rest exhibit power-law indices between $-0.90$ and $-0.27$. Two quasars (J2228+0110 and J2245+0024) do not have enough data, so we did not attempt any spectral model fit to their data.

J1026+2542 and J1146+4037 show evidence of a spectral turnover. The fitted optically thin and thick power-law indices are $-0.58$ and 0.50 in the case of J1026+2542, $-0.90$ and 1.20 in the case of J1146+4037. The turnover frequencies are $\sim234$ MHz ($\sim1.5$ GHz in the rest frame) and $\sim1.8$ GHz ($\sim11.2$ GHz in the rest frame) for J1026+2542 and J1146+4037, respectively. These make these targets GPS/HFP-like sources. The fitted results can be seen in Figure \ref{gcm} and Table \ref{fitr}.

\subsection{Origin of the spectral turnover}

Two targets (J1026+2542 and J1146+4037) of our sample show a possible spectral turnover. The origin of the spectral turnover can be investigated considering the magnetic field and the spectral age. The difficulties are determining the turnover frequency and measuring the source size at the turnover frequency. We found observations at mas resolution  at the frequency near the turnover frequency for J1146+4037  \citep{Frey2010}.

{\bf Magnetic field - }In this work, we consider two methods to estimate the magnetic field strength. An indirect way to estimate the magnetic field of compact radio sources is to assume that the radio emission is in a near equipartition of energy between the radiating particles and the magnetic field \citep{Pacholczyk1970}. Although this condition is assumed in many evolutionary models, there is no a priori reason to believe that magnetic fields in radio sources are in equipartition. 
\citet{Orienti2008, Orienti2012} supported this hypothesis by studying the equipartition magnetic field in a number of HFPs. In this work, we follow the method of these authors by assuming an ellipsoidal geometry with a filling factor of unity by means of the standard formulae in \citet{Pacholczyk1970}, who assume that  proton and electron energies are equal. Then the equipartition magnetic field strength can be determined by Equations 1--3 in \citet{Orienti2012}, who use the 8.4 GHz luminosity and source size, but found a weaker dependence on other measured quantities that were poorly constrained by physical parameters.

A direct way to measure the magnetic field is by means of the spectral parameters for radio sources that have convex spectra. If the spectral peak is produced by SSA, we can compute the magnetic field using observable quantities alone. The main difficulty in adopting this method is the uncertainty in determining source parameters at the turnover frequency, which limits the accuracy of the magnetic field estimate. However, this method may be used for GPS/HFP sources. The peak frequency around a few GHz allows for sampling both the optically thick and  thin parts of the spectrum by multifrequency, high-resolution observations especially for high-redshift sources, leading to a fairly accurate estimate of the peak parameters. Under the SSA assumption, the magnetic field $H$ can be measured directly from the spectral peak parameters, that is, the peak frequency $\nu_{\rm p}$ in units of GHz in the observed frame, the corresponding flux density $S_{\rm p}$ in units of Jy in the observed frame, and the source angular sizes $\theta_{\rm maj}$ and $\theta_{\rm min}$ in units of mas at the turnover \citep{Kellermann1981}. This equation is written as
\begin{equation}
H\sim f(\alpha)^{-5} \theta_{\rm maj}^{2} \theta_{\rm min}^{2} \nu _{\rm p}^{5} S_{\rm p}^{-2} (1 + z)^{-1} .
\label{h}
\end{equation}
We adopt $f(\alpha)\sim8$ ($\alpha = -0.5$), as $f(\alpha)$ weakly depends on $\alpha$ \citep{Kellermann1981}.

In the case of J1026+2542, the spectral turnover appears in the low-frequency region (i.e., $<300$ MHz), which is dominated by the data from the GLEAM survey \citep{Hurley-Walker2017}. These data were observed simultaneously. Thus, here we
may be observing a real turnover, not one that is mimicked by
the blazar variability. As there are no data to constrain the source
size at the peak frequency towards this target, we cannot derive
any information on the magnetic field.

 To derive the equipartition magnetic field of J1146+4037,
we predict the rest-frame 8.4 GHz (redshifted to 1.4 GHz at
$z = 5.0059$) flux density from our spectral model. However, there
is no source size measurement at 1.4 GHz. We make use of the
full width at half maximum (FWHM) source size of  $0.74\pm0.01$
mas derived by the Gaussian fit from 5 GHz VLBI mas angular
resolution observations (Frey et al. 2010).
We note that in our calculations, we assume a source size that is 1.8 times larger than the FWHM, following the approach of \citet{Readhead1994} and \citet{Orienti2008}.  The derived equipartition magnetic field is $34^{+8}_{-7}$ mG. This is within the range of the equipartition magnetic fields of 17 HFP radio sources (7$-$60 mG; quasars and galaxies at $0.22<z<2.91$; \citealt{Orienti2012}) and 5 HFPs at $0.084<z<1.887$ (18$-$160 mG; \citealt{Orienti2008}). The magnetic field calculated from the turnover information listed in Table \ref{fitr} is $1.8^{+2.3}_{-2.7}$ G assuming an SSA origin with Equation \ref{h}, however the uncertainty is very large. The large uncertainty is caused by the fact that we only have four data points to constrain the turnover information and we do not have source size measurements at the turnover frequency, but rather we assume the source size measured at another frequency. More data taken in  other wavelength bands are needed to meaningfully constrain  the turnover peak, and mas resolution  observations at the peak frequency are needed to give reliable magnetic field strength measurements. This may indicate that the turnover is not caused by  SSA, by comparing the large magnetic field strength measured from the spectral turnover ($1.8^{+2.3}_{-2.7}$ G) with the equipartition magnetic field strength ($34^{+8}_{-7}$ mG). As J1146+4037 is a strong blazar, the turnover may be caused by its strong variability.  Another possible explanation for the spectral turnover is that high-density plasma in the nuclear region attenuates the radio emission from the central active BH. High-resolution, interstellar medium observations of the nuclear region of this target may address the latter issue.

{\bf Spectral age - }The age may be the most critical parameter for understanding the nature of these intrinsically compact sources. In the ``young scenario'', the HFPs are interpreted as young radio sources, and they evolve into extended radio sources as the inner region expands where  the SSA is responsible for the curvature of their spectrum.  A relationship between the rest-frame peak frequency and the projected linear size (e.g., \citealt{ODea1997}) indicates that the turnover frequency is related to the source dimension, and thus to the source age.  A method known as ``spectral aging'' relates the curvature of the synchrotron spectrum to the age of the radiating particle (\citealt{Kardashev1962};  \citealt{Jaffe1974}), where synchrotron losses first use up high-energy electron populations, resulting a steepening in the emission spectrum as follows:

\begin{equation}
\label{sa}
t_{\rm syn} = 5.03 \times 10^{4}  H^{-1.5}[(1 + z)\nu _{\rm p} ]^{-0.5} \ \rm yr,
\end{equation}
where the radiative age ($t_{\rm syn}$) is in units of years, the magnetic field ($H$) in units of mG, and the break frequency ($\nu_{\rm p}$) in units of GHz.\ The quantity $z$ is the source redshift \citep{Murgia2003}.

Based on the SSA assumption, the derived radiative lifetime of the electron population is 0.2-year, adopting $H = 1.8$ G for J1146+4037 by Equation \ref{sa}. The estimated timescale is too short if the turnover is due to SSA. The turnover we observed may be due to FFA or source variability. The FFA origin is consistent with the idea that the smallest sources (i.e., HFPs) reside within the innermost region of the host galaxy, characterized by an extremely dense and inhomogeneous ambient medium of high electron density.

In the case of the remaining targets in our sample, we also have future VLA high-frequency observations and the upgraded GMRT low-frequency observations, in order to investigate the possible spectral turnover feather (turnover frequency and source intensity at spectral turnover). 

\subsection{Radio loudness}

Most previous characterizations of an extragalactic radio source as ``radio loud'' were based on 1.4 GHz fluxes and simply assumed  a power law. However, with the new GMRT observations in this work, we can directly measure the spectral indices and extrapolate the flux density at rest-frame 5 GHz, especially for sources that have an increasing power-law radio spectra. 

In our work, following \citet{Stocke1992} and \citet{Kellermann1989} we consider two definitions of radio loudness as follows:
\begin{equation}
\label{rl1}
R_{2500\rm\, \AA} \ = \ \frac{S_{5\rm\, GHz}}{S_{2500\rm\, \AA}},
\end{equation}
\begin{equation}
\label{rl2}
R_{4400\rm\, \AA} \ = \ \frac{S_{5\rm\, GHz}}{S_{4400\rm\, \AA}}, 
\end{equation}
where $S_{2500\rm\, \AA}$, $S_{4400\rm\, \AA}$, and $S_{5\rm\, GHz}$ are the rest-frame $2500\rm\, \AA$, $4400\rm \, \AA,$ and 5 GHz flux density, respectively.

We calculate the 5 GHz flux density using the radio SEDs we determined. In the cases of J2228+0110 and J2245+0024, we did not detect 323 MHz emission and there is only $\sim1.4$ GHz data available. We then predict the 5 GHz flux density based on the 1.4 GHz flux density and assuming a power-law $S_{\nu} \propto \nu^{-0.7}$ distribution. The radio loudness $R_{2500\rm\, \AA}$ of our sample spans a very large range from 12$^{+13}_{-13}$ to 4982$^{+279}_{-254}$ shown in Table \ref{rl}.

\section{Summary} \label{sum}

We report on 323 MHz radio continuum observations by GMRT for 13 radio-loud quasars at $z>5$.
Below are our main results: 

1. Ten quasars are detected above $4\sigma$, two are not detected, while data for another one  is contaminated by RFI. Based on our measurements together with archival data, nine quasars have power-law radio SEDs, and two may have spectral peaks. 

2. Model fitting of the radio SEDs of the sources with power-law behavior shows that the power-law indices have a positive range between 0.18 and 0.67 for some sources and a negative range between  $-0.90$ and $-0.27$ for others. For the two sources with curved spectra, the radio peak centers at around $\sim1$ and $\sim10$ GHz  in the rest frame, the optically thin indices are $-$0.58 and $-$0.90, and the optically thick indices are 0.50 and 1.20. 

3. A magnetic field and spectral age analysis of J1146+4037 may indicate that the turnover is not due to SSA. Source variability or FFA by the ambient plasma in the nuclear region may be the interpretations for the spectral turnover of this target.

4. With the 5 GHz flux density from the spectra modeling, we calculate the radio loudness $R_{2500\rm\, \AA}$, which spans a wide range from 12$^{+13}_{-13}$ to 4982$^{+279}_{-254}$.

\begin{sidewaystable*}
\caption{Sample and archival data I}
\label{pdata1}
\tiny
\centering
\begin{tabular}{lccccccccccc}
\hline\hline
Source&$z$&$S_{120 \rm\, MHz}$ &$S_{147.5 \rm\, MHz}$ &$S_{232 \rm\, MHz}$ &$S_{325 \rm\, MHz}$&$S_{1.4 \rm\, GHz}$&$S_{1.6 \rm\, GHz}$&$S_{1.7 \rm\, GHz}$&$S_{5 \rm\, GHz}$&$S_{8.4 \rm\, GHz}$ &$S_{32 \rm\, GHz}$\\
  &  & (mJy) &(mJy)&(mJy) &(mJy)& (mJy)& (mJy)& (mJy)&(mJy)&(mJy)&(mJy)\\
\hline
J0131$-$0321 &$^{t}5.18\pm0.01$&-&$^{x, 1}25.70\pm4.70$&-&-&$^{w}33.69\pm0.12$&-&$^{v}64.40\pm0.30$&-&-&-\\
J0741+2520&$^{k}$5.194&-&-&-&-&$^{w}2.97\pm0.14$&-&-&-&-&-\\
J0836+0054&$^{e}5.774\pm0.003$&-&-&-&-&$^{d}1.75\pm0.04$&$^{b}1.10\pm0.03$&-&$^{f}0.34\pm0.04$&-&-\\
J0913+5919&$^{m}5.1224\pm0.0001$&-&-&$^{h}30.00\pm3.00$&-&$^{c}19.41\pm0.12$&-&-&$^{d}8.10\pm0.20$&-&-\\
J1034+2033&$^{z}5.0150\pm0.0005$&-&-&-&-&$^{w}3.85\pm0.14$&-&-&-&-&-\\
J1146+4037&$^{m}5.0059\pm0.0007$&-&-&-&-&$^{w}12.53\pm0.15$&$^{l}15.50\pm0.80$&-&$^{l}8.60\pm0.40$&-&-\\
J1427+3312&$^{g}$6.12&-&-&-&$^{u}4.80\pm0.80$&$^{j}1.78\pm0.11$&$^{i}$0.92 &-&$^{i}$0.46 &$^{j}0.25\pm0.02$&-\\
J1429+5447&$^{p}6.1831\pm0.0007$&$^{y}10.86\pm1.55$&-&-&-&$^{w}2.95\pm0.15$&$^{n}3.30\pm0.06$&-&$^{n}0.99\pm0.06$&-&$^{p}0.26\pm0.02$\\
J1614+4640&$^{z}5.3131\pm0.0013$&-&-&-&-&$^{w}2.16\pm0.14$&-&-&-&-&-\\
J2228+0110&$^{q}$5.95&-&-&-&-&$^{o} 0.31\pm0.06$&$^{s}0.30\pm0.12$&-&-&-&-\\
J2239+0030&$^{r}$5.09&-&-&-&-&$^{o}1.22\pm0.08$&-&-&-&-&-\\
J2245+0024&$^{a}$5.17&-&-&-&-&$^{o}1.09\pm0.06$&-&-&-&-&-\\
\hline
\end{tabular}
\tablebib{$^{a}$\citet{Sharp2001}; $^{b}$\citet{Frey2003}; $^{c}$\citet{Momjian2003}; $^{d}$\citet{Petric2003}; $^{e}$\citet{Stern2003}; $^{f}$\citet{Frey2005}; $^{g}$\citet{McGreer2006}; $^{h}$\citet{Carilli2007}; $^{i}$\citet{Frey2008}; $^{j}$\citet{Momjian2008}; $^{k}$\citet{McGreer2009}; $^{l}$\citet{Frey2010}; $^{m}$\citet{Hewett2010}; $^{n}$\citet{Frey2011}; $^{o}$\citet{Hodge2011}; $^{p}$\citet{Wang2011}; $^{q}$\citet{Zeimann2011}; $^{r}$\citet{McGreer2013}; $^{s}$\citet{Cao2014}; $^{t}$\citet{Yi2014}; $^{u}$\citet{Coppejans2015}; $^{v}$\citet{Gabanyi2015}; $^{w}$\citet{Helfand2015}; $^{x}$\citet{Intema2017}; $^{y}$\citet{Shimwell2017}; $^{z}$SDSS}
\tablefoot{Column 1: source name. Column 2: redshift. Columns 3$-$12: the flux density of each observing frequency. \\
\tablefoottext{1}{We pulled the image containing our source from TGSS \citep{Intema2017}, and adopted {\ttfamily{CASA}} 2D Gaussian fit tool to calculate the flux density as well as its error. This source does not appear in the TGSS online dataset because they used a 7$\sigma$ cut.}
}
\end{sidewaystable*}

\begin{table*}
\caption{Sample and archival data II}
\label{pdata2}
\tiny
\centering
\begin{tabular}{lccccc}
\hline\hline
Source&$z$ &Wavelength &$S_{\nu}$&Survey name&Reference\\
 & &(MHz) &(mJy)& & \\
\hline
J1026+2542&$5.2843\pm0.0006$&&&&$k$\\
&&76&-&&\\
&&84&$359.24\pm129.27$&&\\
&&92&$315.76\pm123.45$&&\\
&&99&$419.46\pm111.99$&&\\
&&107&$318.93\pm100.71$&&\\
&&115&$269.68\pm86.41$&&\\
&&122&$507.62\pm67.50$&&\\
&&130&$443.35\pm61.73$&&\\
&&143&$456.76\pm40.78$&&\\
&&151&$400.49\pm35.57$&&\\
&&158&$443.97\pm32.13$&&\\
&&166&$405.07\pm31.00$&GLEAM&$i$\\
&&174&$405.18\pm39.58$&&\\
&&181&$431.84\pm39.13$&&\\
&&189&$422.07\pm39.65$&&\\
&&197&$431.17\pm42.62$&&\\
&&204&$476.86\pm70.88$&&\\
&&212&$527.16\pm71.72$&&\\
&&220&$287.54\pm72.41$&&\\
&&227&$450.06\pm84.39$&&\\
\cline{3-6}
&&147.5&$450.50\pm5.90$&TGSS&$j, 1$\\
&&151&520.00&7C&$e$\\
&&365&$406.00\pm24.00$&Texas&$c$\\
&&408&328.00&B2&$a$\\
&&1400&$239.44\pm0.14$&FIRST&$h$\\ 
&&1700&180.40&-&$g$\\
&&4830&$116.00\pm6.44$&MIT&$b$\\
&&4850&$142.00\pm13.00$&GB6&$d$\\
&&5000&79.20&CLASS&$f$\\
&&8400&105.70&-&$m$\\
&&15000&55 $\pm$ 4&-&$l$\\
&&31000&33 $\pm$ 4&-&$l$\\
&&43000&$55\pm4$&-&$m$\\
&&91000&$14\pm3$&-&$l$\\
\hline
\end{tabular}
\tablebib{$^{a}$\citet{Colla1972}; $^{b}$\citet{Langston1990}; $^{c}$\citet{Douglas1996}; $^{d}$\citet{Gregory1996}; $^{e}$\citet{Waldram1996}; $^{f}$\citet{Myers2003}; $^{g}$\citet{Frey2015}; $^{h}$\citet{Helfand2015}; $^{i}$\citet{Hurley-Walker2017}; $^{j}$\citet{Intema2017}; $^{k}$SDSS; $^{l}$\citet{Sbarrato2013}; $^{m}$\citet{Frey2013}}
\tablefoot{Column 1: source name. Column 2: redshift. Columns 3$-$4: observed wavelength and its corresponding flux density. Columns 5$-$6: survey name and the reference.\\
\tablefoottext{1}{We cut the source image from TGSS database \citep{Intema2017}, and we measured the flux density and the corresponding error with {\ttfamily{CASA}} 2D Gaussian fit tool. The value we present is the same as that in the TGSS online dataset.}
}
\end{table*}

\begin{table*}
\caption{Measurements by GMRT at 323 MHz }
\label{GMRTm}
\tiny
\centering
\begin{tabular}{lcccccccc}
\hline\hline
Source &Phase calibrator&rms &Beam size &$S_{323 \rm\, MHz,\, int}$&$S_{323 \rm\, MHz,\, peak}$ \\
&&(mJy beam$^{-1}$) &(arc sec$^{2}$) &(mJy)&(mJy beam$^{-1}$)\\
\hline
J0131$-$0321&0116--208&-&-&-&-\\
J0741+2520 &0735+331&0.10&$15.74\times6.51$&$1.80\pm0.19$&$1.73\pm0.09$\\
J0836+0054 &0744--064&0.10&$11.07\times7.61$&$2.47\pm0.36$&$1.94\pm0.17$\\
J0913+5919 &0834+555&0.10&$19.57\times6.48$&$9.61\pm0.25$&$9.80\pm0.12$\\
J1026+2542 &1021+219&0.20&$15.62\times6.39$&$318.41\pm0.65$&$317.72\pm0.33$\\
J1034+2033 &3C241&0.10&$10.37\times6.68$&$2.97\pm0.19$&$3.03\pm0.11$\\
J1146+4037 &3C241&0.10&$19.80\times6.14$&$2.98\pm0.15$&$3.05\pm0.07$\\
J1427+3312 &3C286&0.14&$18.09\times6.13$&$4.51\pm0.22$&$3.96\pm0.10$\\
J1429+5447 &3C287&0.17&$26.37\times6.28$&$5.02\pm0.35$&$4.91\pm0.16$\\
J1614+4640 &3C286&0.13&$25.14\times6.12$&$0.81\pm0.19$&$0.94\pm0.09$\\
J2228+0110 &3C454.2&0.15&$10.81\times7.21$&$<0.45^{1}$&-\\
J2239+0030 &2206--185&0.20&$12.78\times6.96$&$1.80\pm0.39$&$1.55\pm0.20$\\
J2245+0024 &3C454.2&0.12&$12.99\times7.23$&$<0.36^{1}$&-\\
\hline
\end{tabular}
\tablefoot{Column 1: source name. Column 2: phase calibrator. Columns 3$-$4: the corresponding rms level and clean beam size for each target  in Figure \ref{GMRT323}. Columns 5$-$6: the integrated and peak flux density measured by {\ttfamily{CASA}} 2D Gaussian tool.\\
\tablefoottext{1}{3$\sigma$ upper limit.}}
\end{table*}

\begin{figure*}
\centering
\subfigure{\includegraphics[scale=0.074]{gmrt323mhz.pdf}} 
\caption{Continuum images from GMRT 323 MHz. The black crosses denote the published optical positions in   \citet{Anderson2001}, \citet{Fan2001}, \citet{Sharp2001}, \citet{McGreer2006, McGreer2009}, \citet{Willott2010}, \citet{Zeimann2011}, \citet{Yi2014}, and SDSS. The shapes of the synthesized beams are plotted in the bottom left corner of each sub-figure. The beam sizes are presented in Table \ref{GMRTm}, where the rms of each map and the flux density of each target are also listed. The contour level, shown by a dashed line, is $-3 \times \rm rms$; the contour levels, shown by a solid line, are $3\times 2^{n} \times \rm rms$, where $n$ is [0, 1, 2, 3, 4......].}
\label{GMRT323}
\end{figure*}

\begin{figure*}
\centering
\subfigure{\includegraphics[scale=0.17]{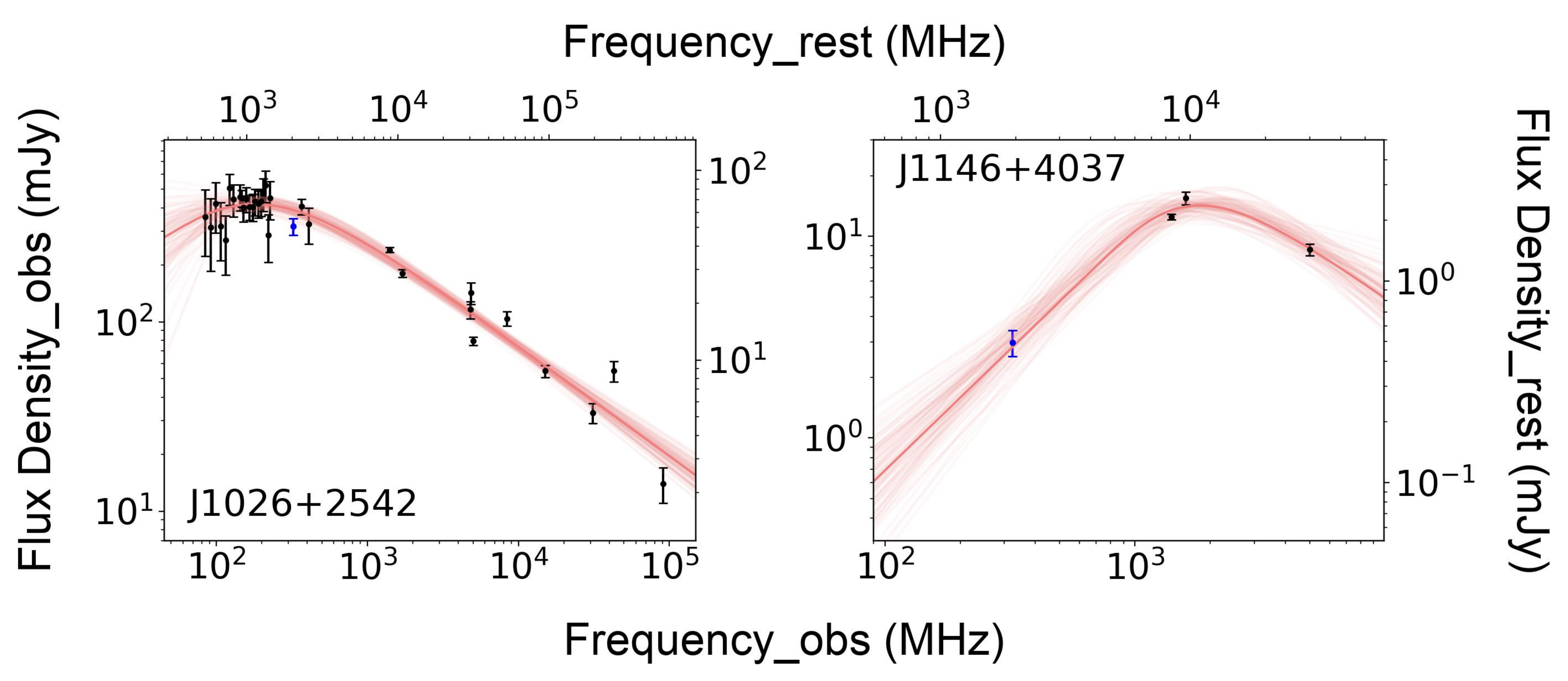}} 
\caption{Spectral model fit with the generic curved model for J1026+2542 (left) and J1146+4037 (right). The description for symbols and lines are same as those in Figure \ref{all}. We present the fitted results in Table \ref{fitr}.}
\label{gcm}
\end{figure*}

\begin{table*}
\caption{Spectral model results}
\label{fitr}
\tiny
\centering
\begin{tabular}{lccccccc}
\hline\hline
&&Power law& &\multicolumn{4}{c}{Generic curved model}\\
\cline{3-3}\cline{5-8}
Source&  &$\alpha_{\rm pl}$&  &$\nu_{\rm p,\, rest}$&$S_{\rm p,\, rest}$&$\alpha_{\rm thin}$&$\alpha_{\rm thick}$\\
 &  & & &(MHz)&(mJy)& & \\
\hline
J0131$-$0321&&$0.29^{+0.11}_{-0.09}$&&-&-&-&-\\
J0741+2520&&$0.34^{+0.14}_{-0.14}$&&-&-&-&-\\
J0836+0054&&$-0.86^{+0.08}_{-0.08}$&&-&-&-&-\\
J0913+5919&&$-0.27^{+0.07}_{-0.05}$&&-&-&-&-\\
J1026+2542&&-&&$1471.0^{+576.6}_{-439.1}$&$64.8^{+4.1}_{-4.7}$&$-0.58^{+0.04}_{-0.03}$&$0.50^{+0.47}_{-0.23}$\\
J1034+2033&&$0.18^{+0.12}_{-0.12}$&&-&-&-&-\\
J1146+4037&&-&&11227.6$^{+2840.8}_{-3266.4}$&2.4$^{+0.2}_{-0.2}$&$-0.90^{+0.30}_{-0.27}$&$1.20^{+0.26}_{-0.20}$\\
J1427+3312&&$-0.90^{+0.05}_{-0.04}$&&-&-&-&-\\
J1429+5447&&$-0.67^{+0.04}_{-0.03}$&&-&-&-&-\\
J1614+4640&&$0.67^{+0.23}_{-0.23}$&&-&-&-&-\\
J2228+0110&&-&&-&-&-&-\\
J2239+0030&&$-0.27^{+0.22}_{-0.22}$&&-&-&-&-\\
J2245+0024&&-&&-&-&-&-\\
\hline
\end{tabular}
\tablefoot{Column 1:  source name. Column 2: the power-law index of the standard nonthermal power-law model. Columns 3$-$6: derived parameters of the generic curved model. }
\end{table*}

\begin{table*}
\caption{Radio loudness}
\label{rl}
\tiny
\centering
\begin{tabular}{lcccccc}
\hline\hline
Source&$m_{z}$&$S_{2500\rm\ \AA}$ &$S_{4400\rm\, \AA}$&$S_{5\rm\, GHz}$&$R_{2500\rm\, \AA}$&$R_{4400\rm\, \AA}$  \\
&(mag)&(mJy) &(mJy)&(mJy) &&\\
\hline
J0131$-$0321&$^{d}$18.08&0.046&0.061&6.71$^{+0.63}_{-0.62}$&147$^{+14}_{-13}$&111$^{+10}_{-10}$\\
J0741+2520&$^{d}$18.44&0.033&0.043&0.40$^{+0.42}_{-0.42}$&12$^{+13}_{-13}$&9$^{+10}_{-10}$\\
J0836+0054&$^{1}$18.83&0.022&0.029&0.33$^{+0.03}_{-0.03}$&15$^{+1}_{-1}$&11$^{+1}_{-1}$\\
J0913+5919&$^{d}$20.81&0.0037&0.0049&2.66$^{+0.22}_{-0.20}$&716$^{+58}_{-53}$&540$^{+44}_{-40}$\\
J1026+2542&$^{d}$19.85&0.0089&0.012&44.21$^{+2.47}_{-2.25}$&4982$^{+279}_{-254}$&3756$^{+210}_{-191}$\\
J1034+2033&$^{d}$19.70&0.010&0.014&0.58$^{+0.50}_{-0.50}$&56$^{+48}_{-48}$&42$^{+36}_{-36}$\\
J1146+4037&$^{d}$19.30&0.015&0.020&1.41$^{+0.34}_{-0.18}$&97$^{+18}_{-13}$&74$^{+14}_{-10}$\\
J1427+3312&$^{1}$18.87&0.021&0.027&0.35$^{+0.03}_{-0.03}$&17$^{+1}_{-1}$&13$^{+1}_{-1}$\\
J1429+5447&$^{b, 1}$21.45&0.0019&0.0025&0.60$^{+0.04}_{-0.04}$&318$^{+19}_{-19}$&240$^{+14}_{-14}$\\
J1614+4640&$^{d}$19.71&0.010&0.013&0.23$^{+0.39}_{-0.39}$&23$^{+39}_{-39}$&17$^{+29}_{-29}$\\
J2228+0110&$^{c, 1}$22.28&0.00089&0.0012&0.07&80&60\\
J2239+0030&$^{d}$21.01&0.0031&0.0041&0.23$^{+0.37}_{-0.37}$&74$^{+121}_{-121}$&56$^{+91}_{-91}$\\
J2245+0024&$^{a}$21.72&0.0016&0.0021&0.26&162&122\\
\hline
\end{tabular}
\tablebib{$^{a}$\citet{Sharp2001}; $^{b}$\citet{Frey2011}; $^{c}$\citet{Zeimann2011}; $^{d}$SDSS.}
\tablefoot{Column 1: source name. Column 2: $z$ band AB magnitude. Columns 3$-$4: rest-frame 2500 $\rm\AA$ and 4400 $\rm\AA$ flux density,which are calculated assuming  a UV power-law $S_{\nu} \propto \nu^{-0.5}$ with $z$-band photometry data. Column 5: rest-frame 5 GHz flux density that is predicted from our spectral model fit. In the case of J2228+0110 and J2245+0024, the 5 GHz flux density is calculated with 1.4 GHz flux density assuming a power-law $S_{\nu} \propto \nu^{-0.7}$. Columns 6$-$7: radio loudness defined in Equation \ref{rl1} and \ref{rl2}.\\
\tablefoottext{1}{$z$-band magnitude is contaminated by the strong Ly$\alpha$ line emission.}
}
\end{table*}

\begin{acknowledgements}
We thank the GMRT staff who helped to observe our project. The GMRT is run by the National Centre for Radio Astrophysics of the Tata Institute of Fundamental Research. R.W. acknowledges supports from the National Science Foundation of China (NSFC) grants No. 11721303, 11991052, and 11533001. D.R. acknowledges support from the National Science Foundation under grant number AST-1614213 and AST-1910107 and from the Alexander von Humboldt Foundation through a Humboldt Research Fellowship for Experienced Researchers.
\end{acknowledgements}

\clearpage
\bibliographystyle{aa} 
\bibliography{mybib}

\end{document}